\documentclass[final]{elsart}

\usepackage[dvips]{graphicx}
\usepackage{amssymb}

\begin{document}

\begin{frontmatter}

\title{Underground Muon Counters as a Tool for Composition Analyses}

\author[Tandar]{A. D. Supanitsky\thanksref{UNAM}\corauthref{cor}},
\author[Tandar]{A. Etchegoyen\thanksref{CONICET}},
\author[UNAM]{G. Medina-Tanco},
\author[CAB]{I. Allekotte},
\author[CAB]{M. G\'omez Berisso\thanksref{CONICET}},
\author[Tandar]{M. C. Medina\thanksref{CONICET2}}
\address[Tandar]{Departamento de F\'isica, Comisi\'on Nacional de Energ\'ia At\'omica, Av. Gral. Paz 1499,
Buenos Aires, Argentina.}
\address[CAB]{Centro At\'omico Bariloche e Instituto Balseiro, CNEA-UNC, (8400) San Carlos de Bariloche, Argentina.}
\address[CONICET]{Member of Carrera del Investigador Cient\'ifico, CONICET, Argentina.}
\address[CONICET2]{CONICET, Argentina.}
\address[UNAM]{Instituto de Ciencias Nucleares, UNAM, Circuito Exterior S/N, Ciudad Universitaria,
M\'exico D. F. 04510, M\'exico.}
\corauth[cor]{Corresponding author. E-mail: supanitsky@nucleares.unam.mx.
Present Address: Instituto de Ciencias Nucleares, UNAM, Circuito
Exterior S/N, Ciudad Universitaria, M\'exico D. F. 04510, M\'exico.}

\begin{abstract}
The transition energy from galactic to extragalactic
cosmic ray sources is still uncertain, but it should be associated either
with the region of the spectrum known as the second knee or with the
ankle. 
The baseline design of the Pierre Auger Observatory was optimized for the
highest energies. The surface array is fully efficient above $3 \times 10^{18}$ eV
and, even if the hybrid mode can extend this range below $10^{18}$ eV, the 
second knee and a considerable portion of the wide ankle structure are left 
outside its operating range. Therefore, in order to encompass these spectral 
features and gain further insight into the cosmic ray composition variation
along the transition region, enhancements to the surface and fluorescence components
of the baseline design are being implemented that will lower the full efficiency 
regime of the Observatory down to $\sim 10^{17}$ eV. 
The surface enhancements consist of a graded infilled area of standard Auger water 
Cherenkov detectors deployed in two triangular grids of 433 m and 
750 m of spacing. Each surface station inside this area will have an associated muon 
counter detector. The fluorescence enhancement, on the other hand, consists of three 
additional fluorescence telescopes with higher elevation angle ($30^\circ-58^\circ$) than the 
ones in operation at present. The aim of this paper is threefold.
We study the effect of the segmentation of the muon counters and
find an analytical expression to correct for the under counting
due to muon pile-up. We also present a detailed method to
reconstruct the muon lateral distribution function for the 750 m spacing array. 
Finally, we study the mass discrimination potential of a new
parameter, the number of muons at 600 m from the shower axis,
obtained by fitting the muon data with the above mentioned 
reconstruction method.

\end{abstract}

\end{frontmatter}

\section{Introduction}
\label{Int}

The cosmic rays energy spectrum extends for about eleven orders of
magnitude, starting at energies below 1 GeV up to energies of
more than $10^{20}$ eV. It presents three main features:
the knee, the second knee and the ankle. There is
evidence of a fourth feature, the so-called GZK suppression
\cite{GZKAuger,GZKHiRes}, which is originated by the interaction
of high energy protons with the photons of the cosmic microwave 
background. In the case of heavier nuclei,
a similar effect is expected due to the fragmentation of
the nuclei in their interaction with the photons of the
microwave and infrared backgrounds \cite{Allard:05a}.

The knee has been observed by several experiments
\cite{Kampert:04,Aglietta:04,Antoni:05} at around $3-5 \times
10^{15}$ eV. At this energy the spectral index changes from $-2.7$
to $-3.1$. The KASCADE data shows that the composition at the knee
presents a transition from light to heavy primaries in such a way
that, at energies above $10^{16}$ eV, the composition is dominated
by heavy nuclei. These particles are originated in our
Galaxy and what is being detected is, very likely, the end of the
efficiency of supernova remnant shock waves as accelerators.

The second knee has been observed at around $4 \times 10^{17}$ eV
by Akeno \cite{Nagano:84}, Fly's Eye stereo \cite{Abu:01}, Yakutsk
\cite{Pravdin:03} and HiRes \cite{HiRes:04}. The physics of this
feature is still unknown, it might be due to the end of the 
efficiency of supernova remnant shock waves as accelerators or a 
change in the diffusion regime in our Galaxy 
\cite{Hoerandel:03,Candia:02}.

The ankle is a broader feature that has been observed by
Fly's Eye \cite{Abu:01} and Haverah Park \cite{Ave:01}, centered 
at approximately the same energy, $\sim 3 \times 10^{18}$ eV. 
These results have been confirmed by Yakutsk \cite{Pravdin:03}, 
HiRes \cite{HiRes:04} and Auger in Hybrid mode \cite{GZKAuger}. 
AGASA also observed the ankle but at higher energy, around 
$10^{19}$ eV \cite{Takeda:03}. As in the case of the second knee, 
the origin of the ankle is still unknown and its physical 
interpretation is intimately related to the nature of the former. 
The ankle could be the transition between the Galactic and 
extragalactic components \cite{Allard:05} or the result of pair 
creation by extragalactic protons in the interaction with the 
cosmic microwave background \cite{Berezinski:04}.

The precise determination of the mean chemical
composition of the cosmic rays in the energy range above 
$\sim 10^{17}$ eV will allow us to understand the origin of the
second knee and the ankle and to know the energy and the speed 
at which the transition between the Galactic and extragalactic
components is given \cite{MedinaTanco:07}. In particular, it will 
permit to decide among the three main models: (i) the mixed composition 
scenario \cite{Allard:05}, in which the composition injected by the 
extragalactic sources is assumed to be similar to the one of the Galactic 
sources and in which the transition takes place in the ankle region, (ii) 
the dip model \cite{Berezinski:04}, in which the ankle is originated by 
the interaction of extragalactic protons with the cosmic microwave 
background and the transition is given at the second knee and (iii) a 
two-component transition from Galactic iron nuclei to extragalactic 
protons, around the ankle energy \cite{Wibig:05}.  

The Pierre Auger Observatory consists of two Observatories situated one 
in each hemisphere. The Southern Observatory, located in Pampa Amarilla 
close to the city of Malarg\"{u}e, Province of Mendoza, Argentina, 
currently consists of nearly 1600 Cherenkov detectors placed in a 1500 m
triangular grid covering an area of 3000 km$^2$ plus four fluorescence 
telescope buildings, with six telescopes each, situated in the 
periphery of the surface array and overlooking it. The construction 
started in 2000 and is going to be completed early in 2008. A complementary 
Northern Observatory will be sited in Colorado, United States of America. 

The Southern Observatory, in its original design, is able to measure cosmic 
rays of energies above $3\times10^{18}$ eV for the surface array and 
$\lesssim 10^{18}$ eV in hybrid mode. Two enhancements, AMIGA 
(Auger Muons and Infill for the Ground Array) \cite{Etchegoyen:07} 
and HEAT (High Elevation Auger Telescopes) \cite{Klages:07}, will 
extend the energy range down to $10^{17}$ eV, encompassing the second 
knee and ankle region where the Galactic-extragalactic transition takes 
place.

AMIGA will consist of 85 pairs of Cherenkov detectors and 30 m$^2$
muon counters buried $\sim 2.5$ m underground, placed in a graded
infill of 433 m and 750 m triangular grids. The AMIGA infill area is 
bound by two hexagons covering areas of 5.9 km$^2$ and 23.5 km$^2$ 
corresponding to the 433 m and 750 m arrays, respectively. The energy 
thresholds of the 433 m and 750 m arrays are $\sim 10^{17}$ eV and 
$\sim 10^{17.6}$ eV, respectively \cite{Medina:06}. On the other hand, 
HEAT will be formed by three additional telescopes of $30^\circ$ to 
$58^\circ$ elevation angle located next to the fluorescence telescopes 
building at Coihueco. They will be used in combination with the existing 
$3^\circ$ to $30^\circ$ elevation angle telescopes at Coihueco as 
well as in hybrid mode with the AMIGA infills.

These enhancements will also allow detailed composition studies 
based on the combined measurement of the atmospheric depth of maximum 
shower development, $X_{max}$, and the shower muon content. These two 
parameters are very sensitive to primary mass composition. Other mass 
sensitive parameters, like the slope of the lateral distribution function, 
rise-time of the signals in the surface detectors, curvature radius, 
etc. strongly depend on them \cite{Chou:05}.

In this paper we will concentrate on the AMIGA muon detectors
\cite{Etchegoyen:07}. These counters will consist of highly
segmented scintillators with optical fibers ending on 64-pixel
multi-anode photomultiplier tubes (PMT). The scintillator strips 
will be equal to those used for the MINOS experiment \cite{MINOS}. 
The current baseline design calls for 400 cm long $\times$ 4.1 cm
wide $\times$ 1.0 cm high strips of extruded polystyrene doped
with fluors, POP ($1\%$) and POPOP ($0.03\%$), and co-extruded 
with TiO$_2$ reflecting coating. They are covered with reflective 
Al foil. To extract the scintillation light, a wavelength shifting 
fiber is glued into a grove which is machined along one face of the 
scintillator strip. A 10 m$^2$ module will consist of 64 
scintillator strips with the fibers ending on an optical connector 
matched to a 64-pixel multi-anode Hamamatsu H7546B PMT of 
2 mm $\times$ 2 mm pixel size, protected by a PVC casing. 
Each muon counter will consist of three of these 64-channel
modules, totalling 192 independent channels covering an
effective area of 30 m$^2$ (actually during the engineering 
array phase one of these 10 m$^2$ modules from each counter 
will be split into two 5 m$^2$ modules for further analyses 
close to the shower core). These muon counters will
be buried alongside a water Cherenkov tank. Each of the
192 channels of the electronics will count pulses above
a given threshold, with an overall counter time resolution 
of 20 ns.

We also present a detailed method for the reconstruction of the
Muon Lateral Distribution Function (MLDF) from data obtained
by the muon counters of the 750 m-array. An associated problem is the 
pile-up effect due to the finite segmentation of the muon
counters. We analyse this problem and propose a correction that
considerably improves the reconstruction of the MLDF. 
The number of muons at 600 m from the shower axis, 
$N_{\mu}^{Rec}(600)$, is extracted from MLDF fits using our 
reconstruction method. Subsequently, the design parameters of the muon 
counters (segmentation and area) are validated by studying the impact of these 
parameters on the total $N_{\mu}^{Rec}(600)$ uncertainty. Finally, 
we study the mass discrimination power of $N_{\mu}^{Rec}(600)$ as 
compared to other parameters normally used in composition analyses:
the maximum development of the longitudinal profile, $X_{max}$, the curvature 
radius of the shower front, $R$, rise time of the signal at the water Cherenkov
detectors, $t_{1/2}$, and slope of the lateral distribution function of the total 
signal deposited in the water Cherenkov detectors, $\beta$. 
Second order effects, like multiple triggering due to electrons scattered
by a single muon, will be dealt with in a subsequent work.

\section{\label{Seg} Muon Counter Segmentation}

The AMIGA counter electronics just counts pulses above a given
threshold, without a detailed study of signal structure or peak 
intensity. This method is very sturdy
since it does not rely on deconvoluting the number of muons from
an integrated signal. It does not depend on the PMT gain or gain
fluctuations nor on the muon hitting position on the scintillator
strip and the corresponding light attenuation along the fiber
track. Neither does it require thick scintillators to control
Poisson fluctuations in the number of photo electrons per
impinging muon. But this one-bit electronics design relies on a
fine counter segmentation to prevent undercounting due to
simultaneous muon arrivals.

\subsection{\label{FirstStSec} Characteristic distance to the nearest station}

The optimum segmentation of the muon counters depends on
the number of muons arriving in the time given by the
system time resolution. Since the MLDF decreases very 
rapidly with the distance to the shower axis we need to 
determine the average position of the closest station,
where pile-up is more significant.

\begin{figure}[!bt]
\begin{center}
\includegraphics[width=13cm]{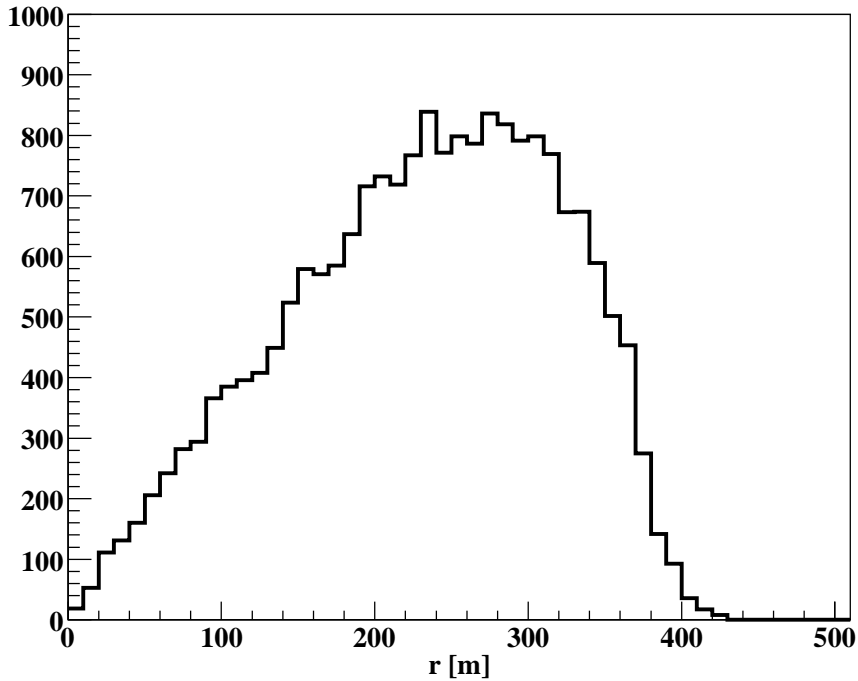}
\caption{Distribution of shower axis to closest station
distance for a $750$ m array for an isotropic incident flux of 
cosmic rays with $\theta \leq 60^\circ$.} 
\label{FirstSt}
\end{center}
\end{figure}

The position of the first station depends on the geometry of the
array and the angular distribution of cosmic rays. We
performed a simple Monte Carlo calculation for the $750$ m-array. 
We uniformly distributed impact points in a triangular 
grid with arrival directions following a $\sin(\theta)
\times \cos(\theta)$ distribution with zenith angle $\theta \leq
60^\circ$ and a uniform distribution for the azimuth angle $\phi$. 
For each event we obtained the distance (along the shower plane) 
between the shower core and the closest station. Figure \ref{FirstSt} 
shows the obtained distribution, with a mean value at $\sim 230$
m. Therefore, we will base our subsequent studies on the number 
of muons found at a characteristic distance of 200 m. 

\subsection{\label{MuonsAtFst} Underground muons at the nearest station}

To prevent contamination due to charged electromagnetic particles
of the shower, the muon counters will be buried underground. Muons 
lose energy mainly by ionization when they propagate through the 
soil. Assuming that the energy loss is proportional to the track 
length, the energy of a muon that travelled a distance $x$ is given 
by,
\begin{equation}\label{EnLoss}
E_\mu (x) = E_{\mu 0} - \alpha \rho x,
\end{equation}
where $E_{\mu 0}$ is the initial energy of the muon, we assume a soil 
density of $\rho = 1.8$ g cm$^{-3}$ and $\alpha = 1.808$ MeV cm$^{2}$ g$^{-1}$ 
is the fractional energy loss per depth of standard rock 
($\langle Z \rangle = 11$ and $\langle A/Z \rangle = 1/2$)
\cite{CrPdg:06}.

Showers initiated by iron nuclei produce more muons than
those from lighter nuclei of the same energy. For that reason
the simulations were performed with iron showers to take into account
the most unfavorable case. We assumed 20 ns of sampling time for the 
system time resolution. 

\begin{figure}[!bt]
\begin{center}
\includegraphics[width=13cm]{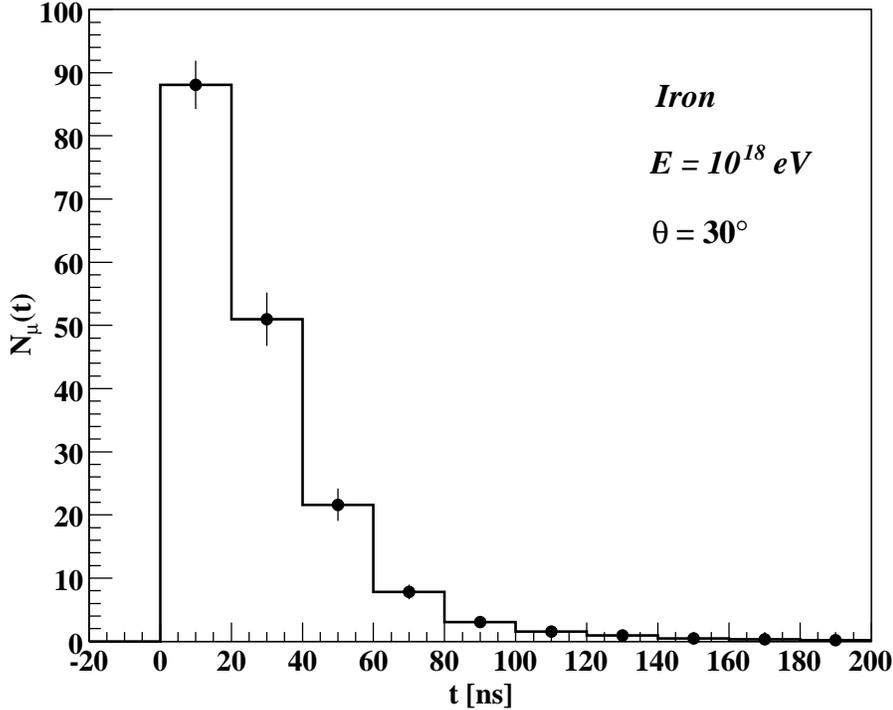}
\caption{Time distribution of muons at $2.5$ m underground and at
200 m from the shower axis in an area of $30 \times \cos(30^\circ)$
m$^2$. The showers were initiated by iron nuclei of $E=10^{18}$ eV
and $\theta=30^\circ$. We used Aires with QGSJET-II to simulate the
50 showers.}
\label{MuonsFst}
\end{center}
\end{figure}

Figure \ref{MuonsFst} shows the time distribution of muons in bins of 
$20$ ns at 200 m from the shower core, in an area of 
30 m$^2$ $\times \cos\theta$ (muon counter area projected on the shower 
plane) and at 2.5 m depth for simulated iron showers of $\theta=30^\circ$, 
primary energy $E=10^{18}$ eV and using QGSJET-II \cite{QGIIa,QGIIb} as the 
high energy hadronic interaction model. We used Aires version 2.8.2 \cite{AIRES} 
to simulate air showers and, to obtain the muon time distribution, we 
propagated them through the soil assuming that they move at the 
speed of light and the energy loss is given by Eq. (\ref{EnLoss}). 
Note that only muons with $E_{\mu 0} > 0.8 \textrm{GeV}/\cos\theta_i$ 
($\theta_i$ zenith angle of the individual muons) reach the detector.

From figure \ref{MuonsFst} we can see that the maximum number of
muons at $E = 10^{18}$ eV and $\theta = 30^\circ$ in the first 20 ns-bin 
is about $90$.

\subsection{\label{Pileup} Muon pile-up correction}

The probability distribution corresponding to a given number of 
muons that hit a segmented detector in a time interval is multinomial,
\begin{equation}
P(n_{1},\ldots,n_{N_{seg}}) = \frac{N_{\mu}!}{n_{1}! \ldots n_{N_{seg}}!}%
\ \frac{1}{(N_{seg})^{N_{\mu}}},
\label{MultiNom}
\end{equation}
where $N_{seg}$ is the number of segments, $N_{\mu}$ is the total number 
of muons and $n_i$ is the number of muons that hit the $i-$th segment, i. e. 
$\sum_{i=1}^{N_{seg}} n_{i} = N_{\mu}$.

As the AMIGA muon counter electronics will be designed to only count
events above a certain threshold, without analysing the corresponding 
signal, whenever two or more muons hit the same scintillator strip in 
the same time bin, these multiple muons will be counted as one. 
Therefore, the total number of muons counted is given by,
\begin{equation}
N_{\mu}^{C}=\sum^{N_{seg}}_{i=1} \Theta(n_{i}),
\label{Ncont}
\end{equation}
where $\Theta(x)=0$ if $x=0$ and $\Theta(x)=1$ if $x\geq1$.

By using Eq. (\ref{MultiNom}) we can calculate the mean value and the variance of $N_{\mu}^{C}$,
\begin{eqnarray}
\label{NcMean} 
<N_{\mu}^{C}>(N_{\mu}) &=& N_{seg} \left[ 1-\left(1-\frac{1}{N_{seg}}\right)^{N_{\mu}}\right], \\
Var\left[N_{\mu}^{C}\right] (N_{\mu}) &=& N_{seg} \left(1-\frac{1}{N_{seg}}\right)^{N_{\mu}}%
\left[1+(N_{seg}-1) \left(1-\frac{1}{N_{seg}-1}\right)^{N_{\mu}} \right. \nonumber \\
&&-N_{seg}\left. \left( 1-\frac{1}{N_{seg}}\right)^{N_{\mu}} \right].
\label{PileUp}
\end{eqnarray}

\begin{figure}[!bt]
\begin{center}
\includegraphics[width=13cm]{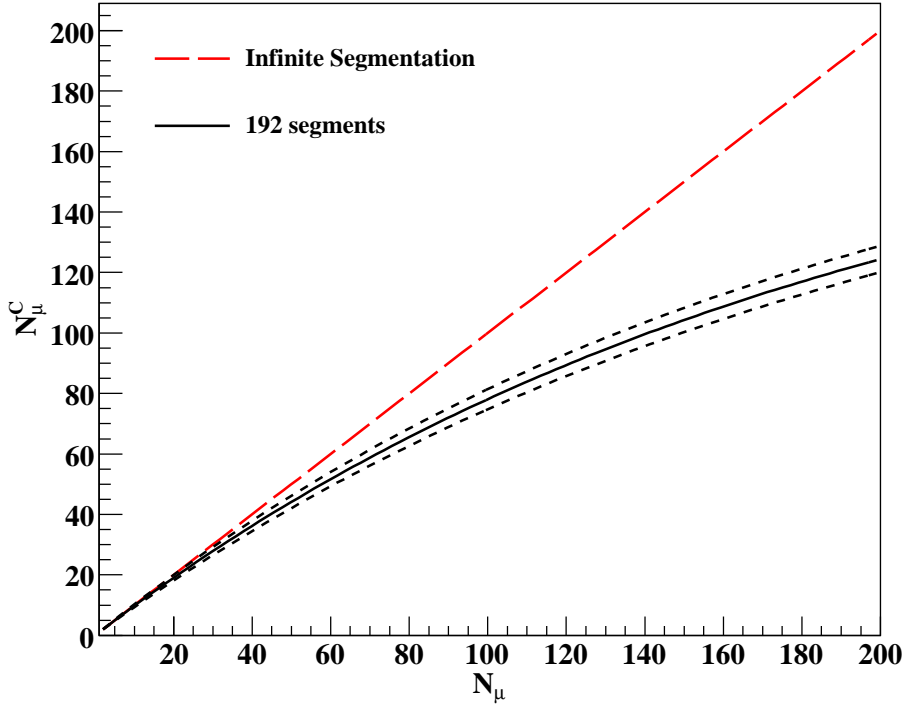}
\caption{Mean value and one sigma region of the number of muons
counted as a function of the incidents for 192 segments (three
PMTs with 64 pixels each). In the ideal case of infinite segmentation
$N_{\mu}^{C}=N_{\mu}$.
\label{NmuC}}
\end{center}
\end{figure}

Figure \ref{NmuC} shows the number of counted muons as a function of 
the incident muons for a muon counter with 192 segments. 

To correct the effect of segmentation we can invert Eq. (\ref{NcMean}),
\begin{equation}
N_{\mu}^{Corr}=\frac{\ln\left(1-\frac{N_{\mu}^{C}}{N_{seg}} \right)}{\ln\left(1-\frac{1}{N_{seg}}%
\right)}.
\label{Corr}
\end{equation}

The number of muons inferred, $N_{\mu}^{Corr}$, obtained from Eq. 
(\ref{Corr}) has an error which can be estimated by solving for a given 
$N_{\mu}^{C}$ the following equations,
\begin{eqnarray}
<N_{\mu}^{C}>(N_{\mu}^{Min})+\sqrt{Var\left[N_{\mu}^{C}\right] (N_{\mu}^{Min})}&=& N_{\mu}^{C}
\label{DNmuMin} \\
<N_{\mu}^{C}>(N_{\mu}^{Max})-\sqrt{Var\left[N_{\mu}^{C}\right] (N_{\mu}^{Max})}&=& N_{\mu}^{C},
\label{DNmuMax}
\end{eqnarray}
where $N_{\mu}^{Min}$ and $N_{\mu}^{Max}$, the unknowns, are
the upper and lower limits of the error interval, respectively.
The error associated with $N_{\mu}^{Corr}$ is asymmetric, this is
due to the shape of $N_\mu^C$ as a function of $N_\mu$. If we
define $\sigma_{+}=N_{\mu}^{Max}-N_{\mu}^{Corr}$ and
$\sigma_{-}=N_{\mu}^{Corr}-N_{\mu}^{Min}$, we can see that
$\sigma_{+}>\sigma_{-}$ and that the difference increases with
$N_{\mu}^{C}$. 

To assess the importance of the error introduced with the 
use of the correction formula, it has to be compared to the 
Poissonian fluctuations inherent to the process of counting 
the muons that hit each scintillator strip in a given time bin.
Therefore, the total error in the determination of the number 
of muons is $\Delta N_{\mu}^{\pm}=\sqrt{\sigma_{\pm}^{2}+\sigma_{Poiss}^{2}}$,
where $\sigma_{Poiss}=\sqrt{N_{\mu}^{Corr}}$ is the error
corresponding to the Poissonian fluctuations.

\begin{figure}[!bt]
\begin{center}
\includegraphics[width=13cm]{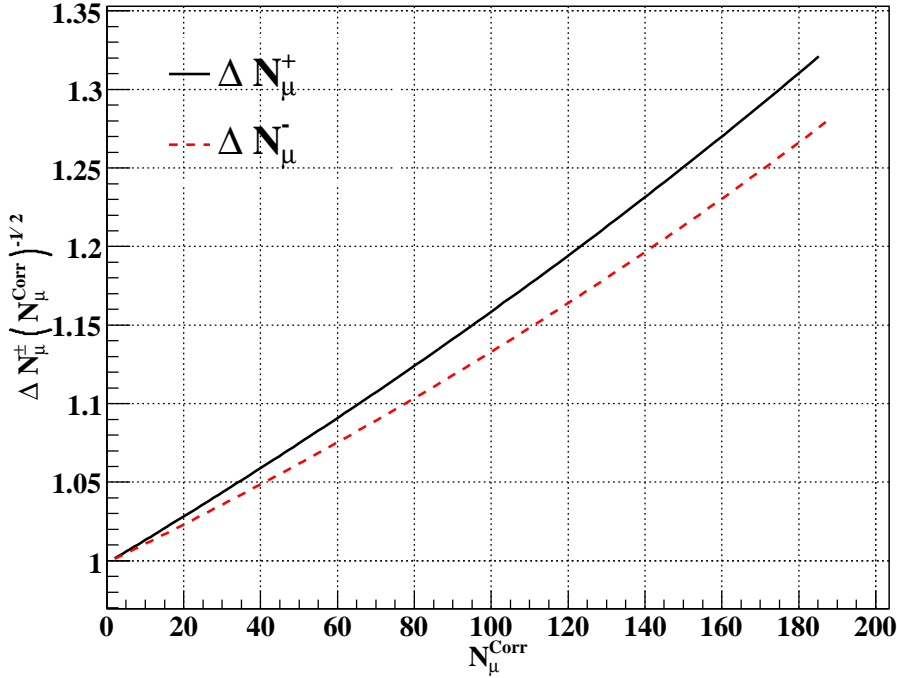}
\caption{Ratio between the total error and the Poissonian error as a
function of the corrected number of muons, for a counter with 192 
segments. The two curves show explicitly the asymmetry of the errors
-see the text for further details.
\label{ErrorSegPois}}
\end{center}
\end{figure}

Figure \ref{ErrorSegPois} shows the ratio between the total error
and the Poissonian one as a function of the number of muons
obtained after correction for 192 segments. From this figure we
see that for 90 incident muons the total error is greater than 
the Poissonian by less than $\sim 14\%$, and as such there is no 
need to further segment the detector. Note that the individual 
uncertainty in each muon counter will not directly translate 
to the $N_{\mu}(600)$ but that it will be further reduced when 
a number of counters are used to fit the MLDF.

It is worth emphasizing that AMIGA envisages an experimental 
verification of the segmentation, since in its unitary cell 
with 7 counters, the modules will have double segmentation.

\section{\label{MLDF} Reconstruction of the MLDF}

One of the first MLDF parametrizations was introduced by
K. Greisen \cite{Greisen:60},
\begin{equation}
\rho_{\mu}(r) = N_{\mu}(t) \left( \frac{r}{r_{0}} \right)^{-0.75} \left( 1+\frac{r}{r_{0}}%
\right)^{-2.5},
\label{LDFMGr}
\end{equation}
where $r$ is the distance to the shower axis, $r_{0} = 320$ m and
$N_{\mu}(t)$ is a normalization constant that depends on the
atmospheric depth $t$. Subsequently other groups proposed
different functional forms \cite{Hillas:70,SUGAR:70}. Although all
these formula describe the MLDF very accurately in the range of 
short and intermediate distances, they are not so good at larger 
distances and for higher-energy showers. Recently the KASCADE-Grande 
Collaboration proposed a new MLDF parametrization which is a modification 
of the Greisen formula \cite{Buren:06},
\begin{equation}
\rho_{\mu}(r) = N_{\mu} \left( \frac{r}{r_{0}} \right)^{-\alpha} \left( 1+\frac{r}{r_{0}}%
\right)^{-\beta_\mu} \left( 1+\left( \frac{r}{10\ r_{0}} \right)^{2} \right)^{-\gamma},
\label{MLDFKascade}
\end{equation}
where $N_{\mu}$, $r_{0}$, $\alpha$, $\beta_\mu$ and $\gamma$ are 
parameters which define the shape and size of the MLDF.

\begin{figure}[!bt]
\begin{center}
\includegraphics[width=14cm]{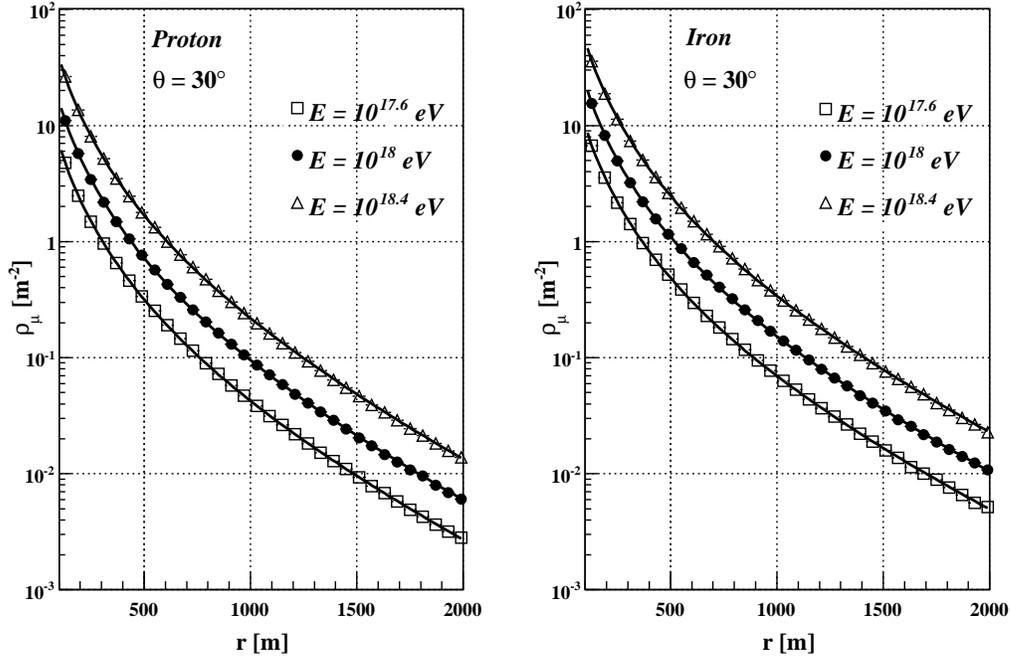}
\caption{Lateral distribution functions of muons at 2.5 m
underground for showers initiated by protons and iron
nuclei of $\theta=30^\circ$ and primary energies of $10^{17.6}$, 
$10^{18}$, and $10^{18.4}$ eV. To generate the air showers we used 
Aires 2.8.2 with QGSJET-II. The solid lines correspond to fits with 
the KASCADE-Grande MLDF. \label{MLDFETH}}
\end{center}
\end{figure}

To study the shape of the underground MLDF we performed air shower 
simulations. Again, we used Aires 2.8.2 with QGSJET-II 
and to propagate the muons in the soil we used Eq. (\ref{EnLoss}). Figure \ref{MLDFETH} 
shows the simulated MLDFs at 2.5 m underground corresponding to showers initiated 
by protons and iron nuclei of $30^\circ$ of zenith angle and for several primary energies. 
It also shows the fits of the profiles with the KASCADE-Grande MLDF where we fixed the 
parameters $r_0 = 320$ m and $\alpha = 0.75$ and left $N_{\mu}$, $\beta_\mu$ and
$\gamma$ as free fit parameters. The simulated profiles are fitted very accurately in 
the range of distances considered.

The range of primary energies for which we
simulated MLDFs in order to fit them with the KASCADE-Grande
formula was $10^{17.6}$ eV - $10^{18.5}$ eV in steps of $\Delta
\log(E/eV)=0.1$. This procedure was performed for both protons and
iron nuclei at $\theta=30^\circ$ and $\theta=45^\circ$. 

The parameter $\gamma$ in Eq. (\ref{MLDFKascade}) describes the behavior of the 
MLDF at relatively large distances from the core, where only low statistics are
available in general. On the other hand, $\beta_\mu$ is well sampled by the 
proposed 750 m-array.

Therefore, for all practical purposes, the analysis shows 
that a better fit of the muon counters data to the complete MLDF function is 
obtained by just fixing $\gamma = 3$, its average value for protons, while leaving 
$\beta_\mu$ and $N_{\mu}$ as free parameters. The factor involving $\gamma$ only 
becomes important at distances to the shower core larger than 
$\sim 10 \times r_0 = 3.2$ km, well beyond the point where data is available at 
the energies of interest. To keep $\gamma$ free when dealing with real data would 
only add degrees of freedom to the fitting function without the corresponding 
increase in the available data.

To develop the reconstruction method it is necessary to simulate an
array of Cherenkov detectors with associated muon counters. The
Cherenkov detectors give the trigger information, geometry and 
energy reconstructions, synchronization, and telecommunications. 
For this purpose we interfaced an ad-hoc developed muon counter 
computer code with the program SDSim version v3r0 \cite{SDSim}, 
which simulates the Pierre Auger surface detectors. 

Since the number of particles produced in a shower is extremely 
large, $\sim 10^{9}$ in a $10^{18.5}$ eV shower, it is practically 
impossible to follow and store all the information of the secondary
particles. Therefore, a statistical method called 
thinning, first introduced by M. Hillas \cite{Hillas:85,Hillas:97}, is
used. An unthinning method has to be applied to calculate the real number 
of muons arriving at a counter. We employed the method introduced by 
P. Billoir described in references \cite{Billoir:00,TesisSupa}.

The muons with enough energy to reach the detector that
arrive in a time interval $\Delta t$ are uniformly distributed and
those that fall into the same segment are counted as one, in this
way we include in the simulation the pile-up effect introduced by
the electronics. The present simulation does not include the
propagation of the electromagnetic particles into the soil.
Detailed numerical simulations involving Geant4 \cite{Geant4}
and Aires show that, beyond $17$ radiation lengths, the 
electromagnetic contribution is at most a few percent \cite{Batata1:07}
and can therefore be neglected in the present analysis. Nevertheless,
these punch through simulations will be experimentally verified 
by a specific detector \cite{Batata2:07}.  

We used the generated showers to study the shape of the MLDF to 
simulate the response of the Cherenkov detectors and muon counters 
by randomly distributing impact points in the 750 m-array. We used 
muon counters of $30$ m$^2$, 192 segments and a time resolution of
$20$ ns. For each muon counter we calculated the ``real'' muon 
distribution (i.e., assuming an infinite segmentation, without 
pile-up effect), the ``measured'' muon distribution (taking into 
account the pile-up effect) and the corrected muon distribution 
(obtained applying Eq. (\ref{Corr}) in each time bin). 

\begin{figure}[!bt]
\begin{center}
\includegraphics[width=13cm]{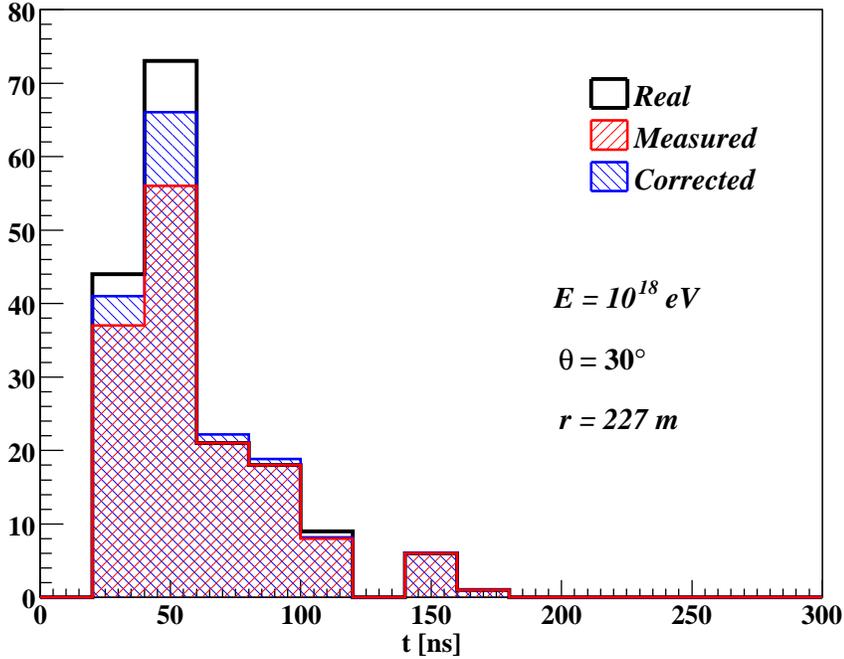}
\caption{Real, measured and corrected time distribution of muons
for the closest station corresponding to one simulated event of
$E=10^{18}$ eV and $\theta=30^{\circ}$. We considered muon
detectors of 30 m$^2$ of area, 192 segments and a time resolution
of 20 ns.}
\label{TDSimEvent}
\end{center}
\end{figure}

As an example, figure \ref{TDSimEvent} shows the ``measured'' time 
distribution of muons in a muon detector at 227 m from the shower 
core for a simulated event initiated by an iron primary of 
$E=10^{18}$ eV and $\theta=30^{\circ}$. Figure \ref{TDSimEvent} 
also shows the ``real'' time distribution and the corrected one.

In this example, the total number of muons corresponding to the 
real distribution is $n_{\mu}^{real} = 172$ and the measured one 
is $n_{\mu}^{med} = 147$, $15\%$ less approximately. The total 
number of muons after applying the correction is 
$n_{\mu}^{corr} = 163$, about $5\%$ less than the real value. 
This shows the importance of the correction in the determination 
of the number of muons at each counter, especially at those close 
to the shower core.

Due to the pile-up effect, counters are not able to
measure the number of muons very close to the shower axis.
Therefore and to avoid systematic uncertainties it is convenient
to fit the MLDFs including stations with a large number of muons in a
different way. We will consider as saturated stations those with
$N_{\mu}^{C} > 72$ (i.e. $N_{\mu}^{Corr} > 90$) in at least one time 
bin.

The probability of having more than 3 background triggers due to 
random coincidences in a time window $\Delta t$ is, 
$P(n\geq3)=1-\exp(-\mu)(1+\mu+\mu^2/2)$, where $\mu=I_0\ A_{mc}\ \Delta t$,
$I_0$ is the vertical intensity of the background and $A_{mc}$ is the muon 
counters area. The background flux at $\sim 2.5$ m underground in the
site where the muon counters are going to be installed is under study.
However, if we assume a vertical intensity like the corresponding to the 
Auger tanks (see Ref. \cite{Bertou:06}) we obtain $P(n\geq3)\sim8\times10^{-7}$ 
and for a vertical intensity 10 times larger, which can be considered as an
upper limit, $P(n\geq3)\sim7\times10^{-4}$.
Therefore, stations with 0, 1 or 2 muons will be considered as silent 
stations to prevent errors coming from such random coincidences. We will 
take into account all silent stations with distance to the shower core less 
than $5000$ m.

The number of muons that hit a given detector follows a Poisson 
distribution. Therefore, to fit the MLDF we minimize the likelihood 
function $\mathcal{L}=-\ln(P)$ with respect to the parameters 
$\vec{p}=(N_{\mu},\beta_\mu)$ of the KASCADE-Grande MLDF (see Eq. 
(\ref{MLDFKascade})), where
\begin{eqnarray}
P &=&
\prod_{i=1}^{N_{sat}} \frac{1}{2} \left( 1-\textrm{Erf}\left( \frac{n_{\mu i}^{meas}-%
\rho_{\mu}(r_i;\vec{p})}{ \sqrt{2 \ \rho_{\mu}(r_i;\vec{p})}} \right) \right)\times \nonumber \\
&& \prod_{i=1}^{N} \exp(-\rho_{\mu}(r_i;\vec{p})) %
\frac{\rho_{\mu}(r_i;\vec{p})^{n_{\mu i}^{corr}}}{n_{\mu i}^{corr}!} \times \nonumber \\
&& \prod_{i=1}^{N_{sil}} \exp(-\rho_{\mu}(r_i;\vec{p})) \left(1+%
\rho_{\mu}(r_i;\vec{p})+\frac{1}{2} \rho_{\mu}(r_i;\vec{p})^2 \right).
\label{Lik}
\end{eqnarray}
Here the first factor corresponds to saturated stations,
the second to ``good'' stations (neither saturated nor
silent) and the third to silent ones. $r_i$ is the
distance of the $i$-th station to the shower core, $N_{sat}$ is
the number of saturated stations, $N$ is the number of ``good''
stations, $N_{sil}$ is the number of silent stations, $n_{\mu
i}^{meas}$  is the total number of muons measured corresponding to
the $i$-th station and $n_{\mu i}^{corr}$ is the total number of
muons for the $i$-th station after applying the correction in
every time bin of the measured time distribution.

The expression for the saturated stations comes from the fact that
the Poisson distribution of mean value $\mu$ can be approximated
by a Gaussian distribution of mean value $\mu$ and
$\sigma=\sqrt{\mu}$, $P(n;\mu) \cong \exp(-(n-\mu)^{2}/2
\mu)/\sqrt{2 \pi \mu}$. Therefore, the probability that $n$ be
greater than a given $n_0$ is,
\begin{equation}
P(n \geq n_{0}) = \frac{1}{\sqrt{2 \pi \mu}} \int^{\infty}_{n_{0}} dn%
\exp\left( -\frac{(n-\mu)^{2}}{2 \mu} \right) =%
\frac{1}{2} \left[ 1-\textrm{Erf}\left( \frac{n_{0}-\mu}{ \sqrt{2 \ \mu}} \right) \right],
\label{Sat}
\end{equation}
where,
\begin{equation}
\textrm{Erf}(x) = \frac{2}{\sqrt{\pi}} \int^{x}_{0} dt \exp(-t^2).
\label{Erf}
\end{equation}

The expression for the silent stations corresponds to the probability that the 
number of muons be less or equal than two,
\begin{equation}
P(n \leq 2) = \sum^{2}_{n=0} \exp(-\mu)\ \frac{\mu^{n}}{n!} = \exp(-\mu) (1+\mu+%
\frac{\mu^{2}}{2}).
\label{Silent}
\end{equation}

\begin{figure}[!bt]
\begin{center}
\includegraphics[width=13cm]{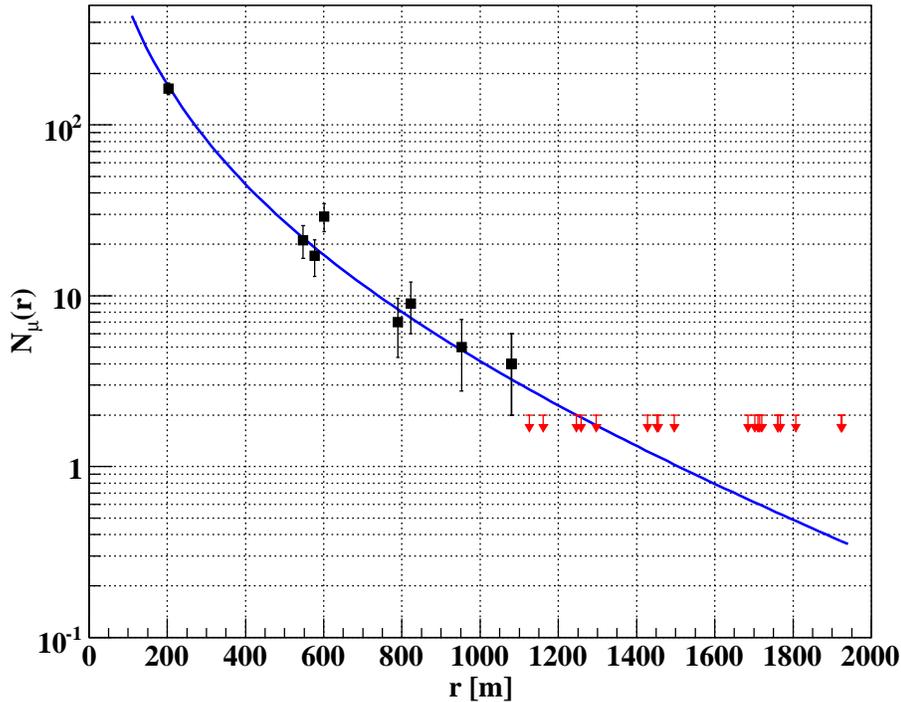}
\caption{Fit of the MLDF for the same event considered in Fig. \ref{TDSimEvent}. 
The arrows indicates the position of the silent stations. We considered muon 
detectors of 30 m$^2$ of area, 192 segments and a time
resolution of 20 ns. \label{Ldf_e18_fe_30deg}}
\end{center}
\end{figure}

Figure \ref{Ldf_e18_fe_30deg} shows a fit of the MLDF for the same
event of figure \ref{TDSimEvent}. To reconstruct the arrival
direction and the position of the shower core we used the standard
package, called CDAS \cite{CDAS}, used to reconstruct the 
information from the Auger Cherenkov detectors. The arrows 
correspond to the positions of the silent stations.

\begin{figure}[!bt]
\begin{center}
\includegraphics[width=10cm]{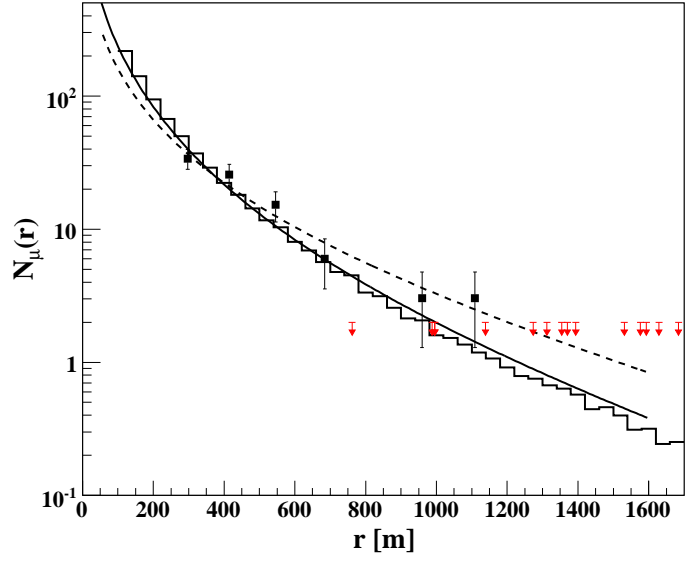}
\includegraphics[width=10cm]{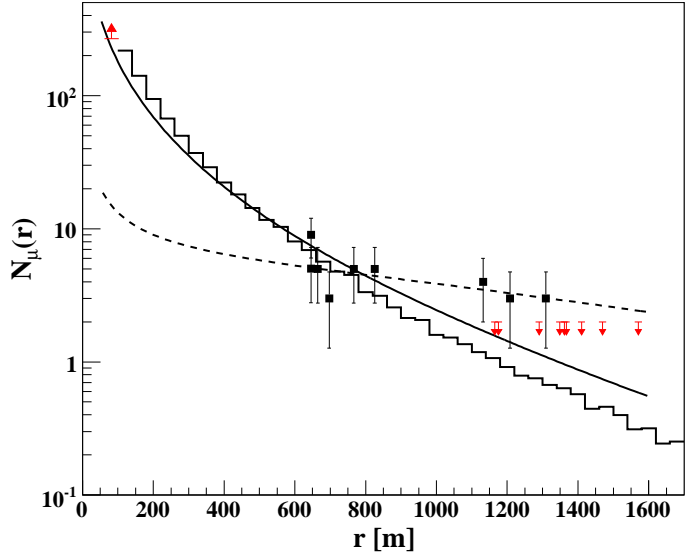}
\caption{Fit of the MLDFs for a proton shower of $E=10^{18}$ eV and $\theta =30^{\circ}$
injected in two different positions of the array. We considered muon detectors
of 30 m$^2$ of area, 192 segments and a time resolution of 20 ns. The arrows indicate 
the position of the saturated (pointing upwards) and silent (pointing downwards) 
stations. The solid blue line corresponds to the fit with saturated and silent stations 
and the dashed black line corresponds the fit without saturated and silent stations. The 
event of the upper panel does not have saturated stations and the one of the bottom panel 
has both saturated and silent stations.
\label{Ldfs_e18_pr_30deg}}
\end{center}
\end{figure}

In order to stress the importance of including the silent and saturated stations
in the fit procedure, figure \ref{Ldfs_e18_pr_30deg} shows the fits corresponding
to proton events of $E=10^{18}$ eV and $\theta =30^{\circ}$, with (solid line) and
without (dashed line) taking into account the silent and saturated stations. We observe
that the fitted MLDF is much more similar, specially close to 600 m from the shower
axis, to the real one (histogram of the figure) for the case in which the silent and
saturated stations are included in the fit. The exclusion of the silent and saturated
stations produces a spurious flattening of the fitted MLDF, specially for those cases in 
which the closest station is saturated, as clearly shown in bottom panel of figure 
\ref{Ldfs_e18_pr_30deg}. The discrepancies between the fitted MLDF obtained without 
including the silent and saturated stations and the real one increases for decreasing 
primary energies. It introduces important biases in particular in the $N_{\mu}(600)$ 
distribution. The inclusion of the saturated and silent stations avoid these biases and 
reduce the error in the determination of $N_{\mu}(600)$.

\section{Muons at 600 m from the Shower Core \label{Nmu600sec}}

The number of muons at a given distance from the shower axis has
been used in the past as a parameter for composition analysis.
The AMIGA 750 m infill has a detector spacing appropriate
to evaluate the number of muons at $600$ m from the shower core
\cite{Hayashida:95}.

The discrimination power of $N_\mu(600)$ depends strongly on its 
reconstruction uncertainty. Therefore, to study the uncertainty 
introduced by the reconstruction method we define,
\begin{equation}
\epsilon = \frac{N_{\mu}^{Rec}(600)}{N_{\mu}^{Real}(600)}-1,
\label{EpsilonDef}
\end{equation}
where $N_{\mu}^{Rec}(600)$ is the reconstructed number of
muons from a MLDF fit and $N_{\mu}^{Real}(600)$ is the expected 
average number of muons from sampling muons in a 40 m wide ring 
(in the shower plane), both at 600 m from the shower axis.

\begin{figure}[!bt]
\begin{center}
\includegraphics[width=6.8cm]{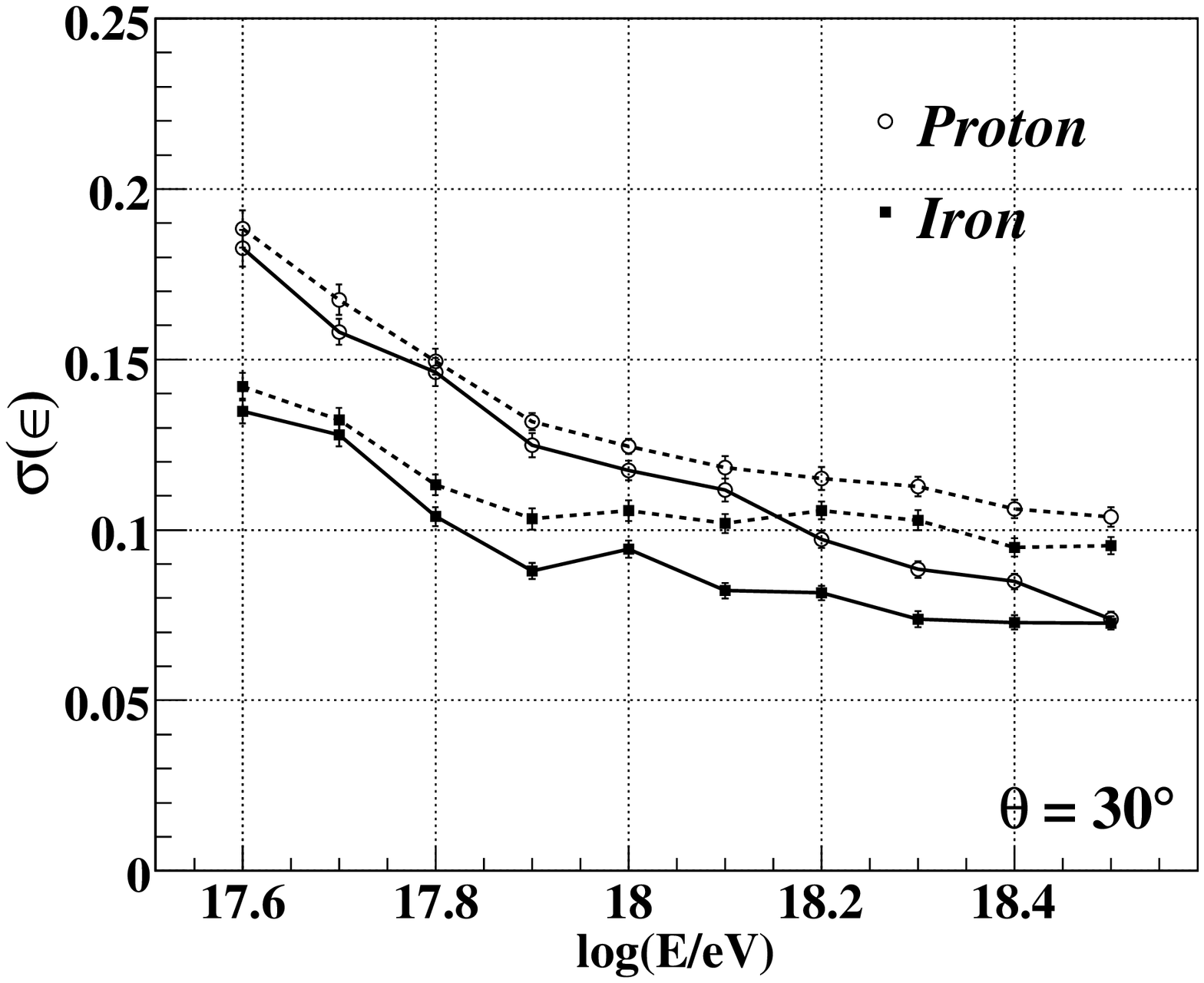}
\includegraphics[width=6.8cm]{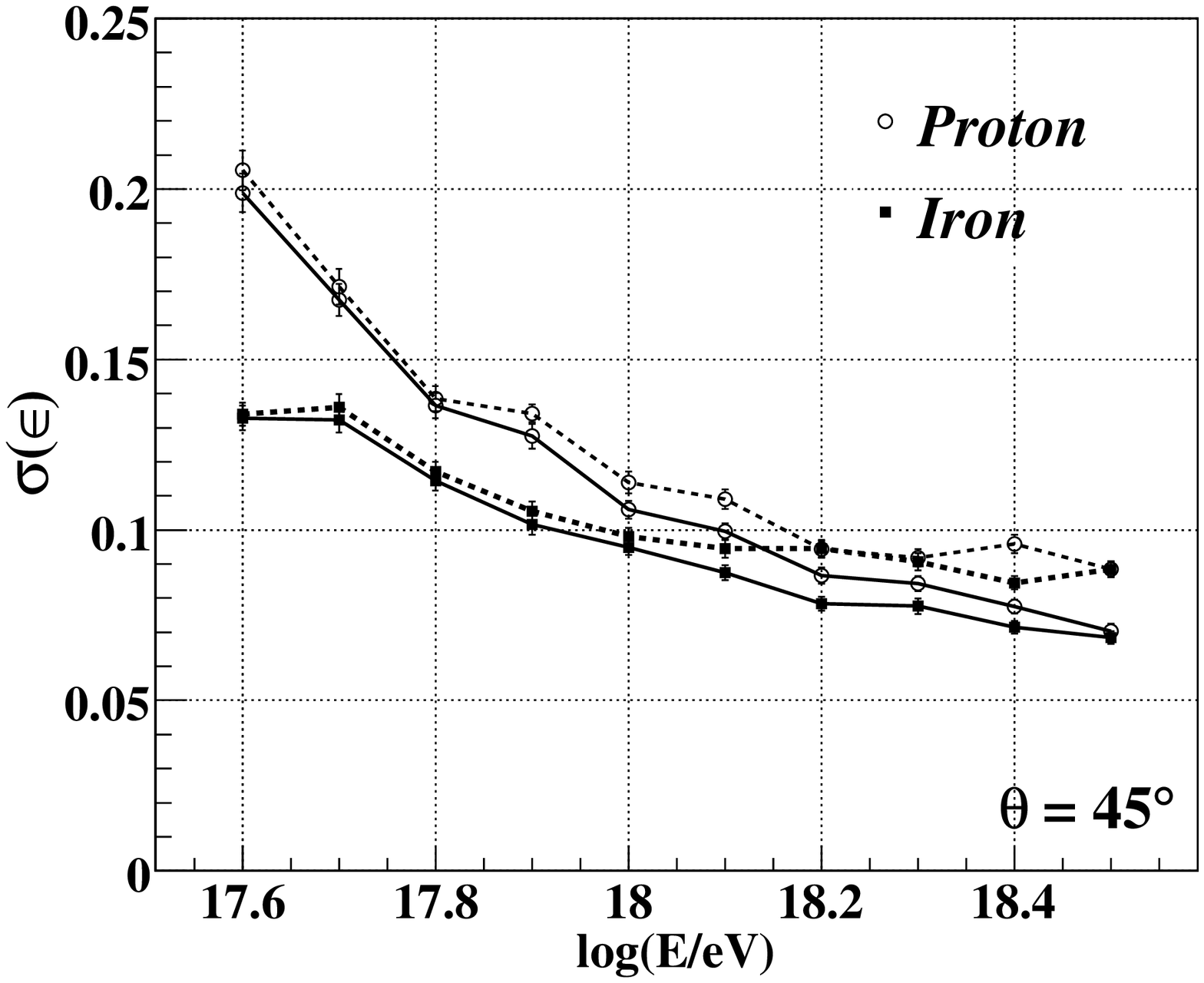}
\caption{Relative error of $N_{\mu}^{Rec}(600)$ as a function of
the logarithm of the energy for protons and iron nuclei of $30^\circ$ 
(left panel) and $45^\circ$ (right panel). We considered muon detectors 
of 30 m$^2$ of area and 192 segments. We assumed a time resolution of 20 ns 
(solid lines) and a value much greater than the width of the time
distribution of muons (dashed lines) which corresponds to the most 
unfavorable case.    
\label{Epsilon}}
\end{center}
\end{figure}

Figure \ref{Epsilon} shows $\sigma(\epsilon)$ (relative error of
$N_{\mu}^{Rec}(600)$), obtained from Gaussian fits of the
distributions of $\epsilon$, as a function of the logarithm of the
energy for proton and iron primaries and zenith angles of
$30^\circ$ and $45^\circ$. We considered muon counters of $30$
m$^2$, 192 segments and a time resolution of 20 ns (solid lines). 
As expected, the relative error decreases with energy, due to the 
fact that the number of muons in the showers increases almost 
linearly with primary energy, therefore, the number of triggered 
muon detectors (with $N_\mu^C \geq 3$) also increases.

From figure \ref{Epsilon} we can also see that the relative error
for the iron nuclei is smaller than the corresponding one for
protons because heavier nuclei produce more muons.

To study the robustness of the reconstruction method we assumed a
time binning much larger than the width of the time distribution of 
muons, the worst case. Figure \ref{Epsilon} shows the obtained results
(dashed lines), the relative error does not increase much, $\lesssim 3\%$ 
with respect to the case of 20 ns.
%
%

Note that for the two cases considered the mean value of $\epsilon$,
i. e. the bias in the determination of $N_\mu (600)$, kept less than 
$\sim 3\%$ in the whole energy range considered (from 
$10^{17.6}$ eV to $10^{18.5}$ eV). 

The uncertainty of $N_{\mu}^{Rec}(600)$ is less than $20\%$ in the
energy range in which the 750 m-array will be effective, from 
$10^{17.6}$ eV to $10^{18.5}$ eV. Moreover, it decreases with energy,
such that for energies greater than $10^{18}$ eV it is smaller than
$13\%$. The error in the reconstruction of the primary energy is
typically around $20\%$ \cite{Roth:07}, and because the number of
muons increases almost linearly with the energy this uncertainty
will dominate in the determination of $N_{\mu}^{Rec}(600)$.
Therefore, muon counters of 30 m$^2$ and 192 segments are
acceptable values for the design parameters. 

\section{Parameters Sensitive to the Chemical Composition}

Several parameters obtained from the surface and fluorescence
detectors are used for the identification of the primary. The
difference between parameters corresponding to different primaries
is due to the fact that showers initiated by heavier primaries
develop earlier and faster in the atmosphere and also have a
larger muon content.

\begin{figure}[!bt]
\begin{center}
\includegraphics[width=15cm]{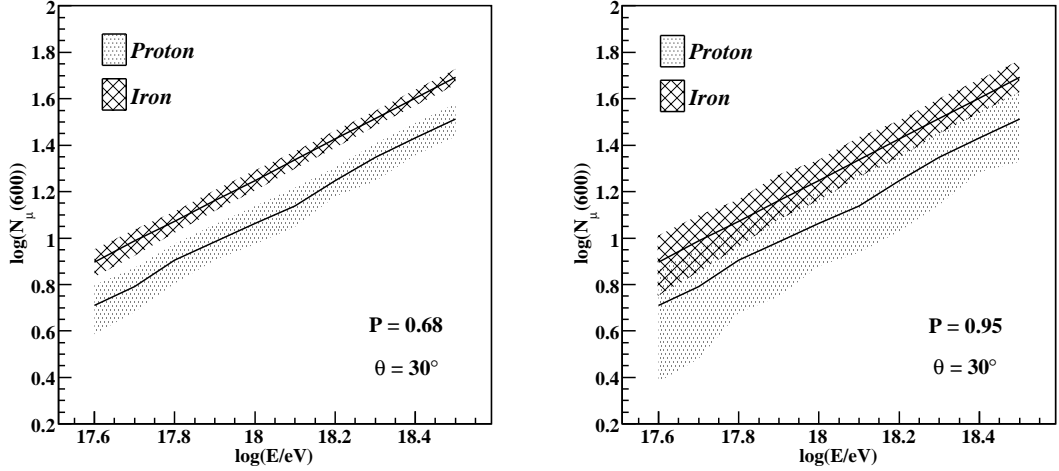}
\caption{$\log(N_{\mu}^{Rec}(600))$ as a function of
$\log(E/\textrm{eV})$ for protons and iron nuclei of
$\theta=30^\circ$. $P$ gives the probability that an 
event falls in the corresponding shadowed region.
\label{Nmu600}}
\end{center}
\end{figure}

From the distributions of $N_{\mu}^{Rec}(600)$ corresponding 
to a given energy, type of primary and zenith angle, we 
calculated the mean value $\langle N_{\mu}^{Rec}(600) \rangle$ 
and the regions of $68\%$ and $95\%$ of probability. 
Figure \ref{Nmu600} shows $\langle N_{\mu}^{Rec}(600) \rangle$ 
and the regions of $68\%$ and $95\%$ probability (parameter P 
in the figure) as a function of the energy for protons and 
iron nuclei of 30$^\circ$ of zenith angle. Although the distributions 
for iron and proton show an overlap, the separation is good at $68\%$ 
confidence level. 

Besides the muon content of the shower and the depth of the
maximum there are other parameters that are used for composition
analysis \cite{supa:05}. These are the structure of the
shower front (in particular the rise-time of the signals in the
surface detectors), the radius of curvature, $R$, and the slope, 
$\beta$, of the lateral distribution function of the total signal 
deposited in the water Cherenkov detectors.

For any given event, we can define a parameter related to the shower
front structure, using the rise-time of the signals in a selected subset of 
the triggered water
Cherenkov detectors,
\begin{equation}
t_{1/2} = \frac{1}{N_{T}} \sum^{N_{T}}_{i=1} (t^{i}_{50}-t^{i}_{10}) \times \left(%
\frac{400\ \textrm{m}}{r_{i}} \right)^{2},
\label{RiseTimeDef}
\end{equation}
where $N_{T}$ is the number of stations with signal greater than 
$10$ VEM \footnote{Vertical Equivalent Muon, signal deposited in a 
water Cherenkov tank when fully traversed by a muon vertically 
impinging in the center of the tank \cite{Bertou:06}}, $t^{i}_{10}$ 
and $t^{i}_{50}$ are the times at which $10\%$ and $50\%$ of the 
total signal is collected, respectively, and $r_{i}$ is the distance 
of the $i$-th station to the shower axis. Only stations at a distance 
to the shower axis greater than $400$ m are included in Eq. 
(\ref{RiseTimeDef}).

To obtain $X_{max}$, including the effect of the detectors and the
reconstruction procedure, we assumed that the distribution of the
reconstructed $X_{max}$ is a Gaussian with an energy dependent 
$\sigma$ as given in Ref. \cite{Klages:07}. Therefore, we obtained 
the distributions of the reconstructed $X_{max}$ from the simulated 
showers by sampling a value from a Gaussian distribution with the 
mean value given by $X_{max}$ calculated internally in Aires and 
$\sigma$ from the interpolation of the simulated data of
Ref. \cite{Klages:07}.

To compare the discrimination power of the different parameters
considered ($q=N_\mu(600), X_{max}, t_{1/2}, \beta, R, S_{600}$) 
we define,
\begin{equation}
\eta(q) = \frac{\left| \langle q_{pr} \rangle - \langle q_{Fe} \rangle \right|}{\sqrt{%
\sigma^{2}(q_{pr}) + \sigma^{2}(q_{Fe})}},
\label{EtaDef}
\end{equation}
where $q_{A}$ is the parameter for the nucleus $A$, $\langle
q_{A} \rangle$ is the mean value and $\sigma(q_{A})$ the standard
deviation for the distribution of $q_{A}$. From the definition of
$\eta(q)$ we see that the larger its value, the greater the 
discrimination power of the parameter $q$.

\begin{figure}[!bt]
\begin{center}
\includegraphics[width=6.8cm]{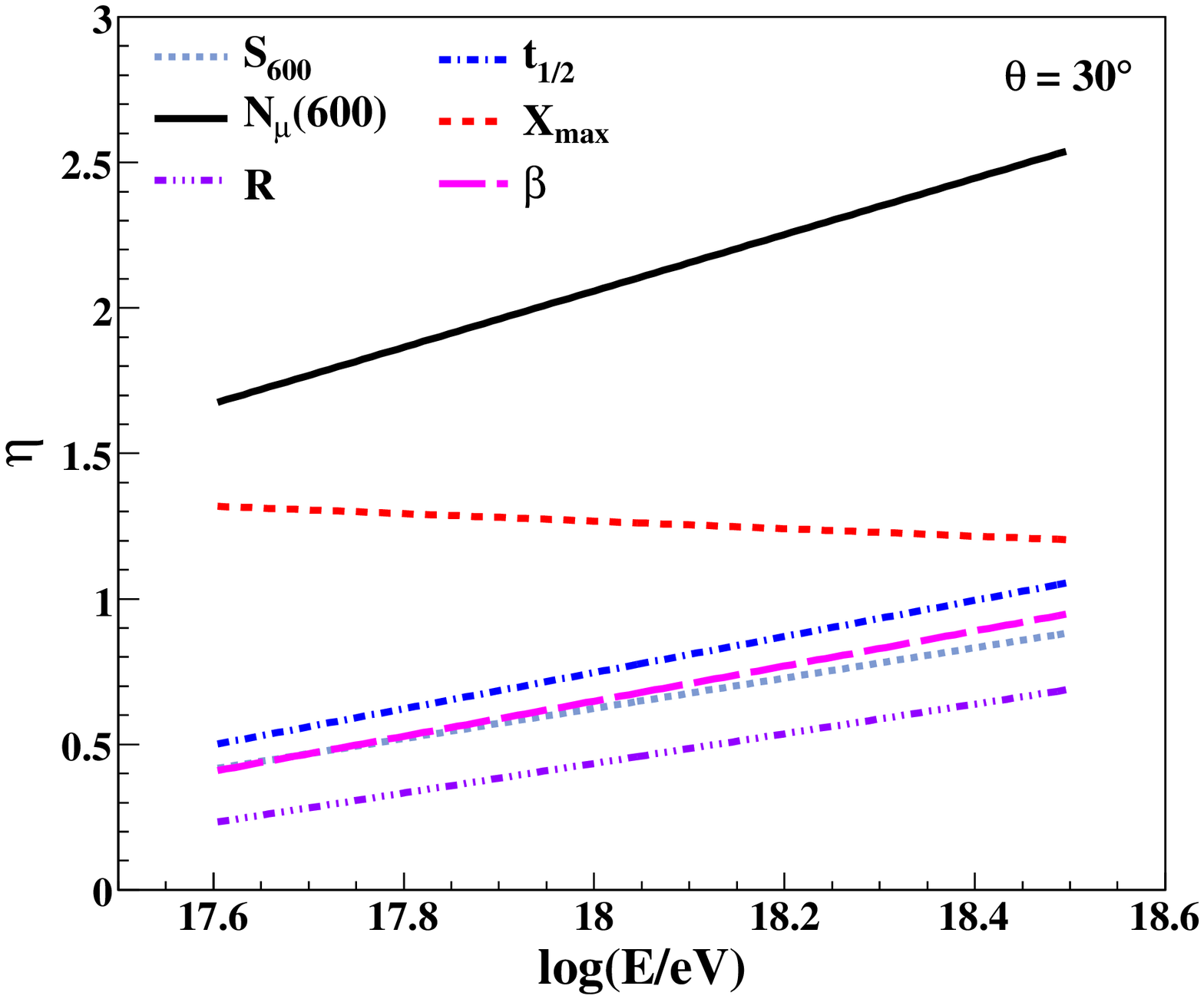}
\includegraphics[width=6.8cm]{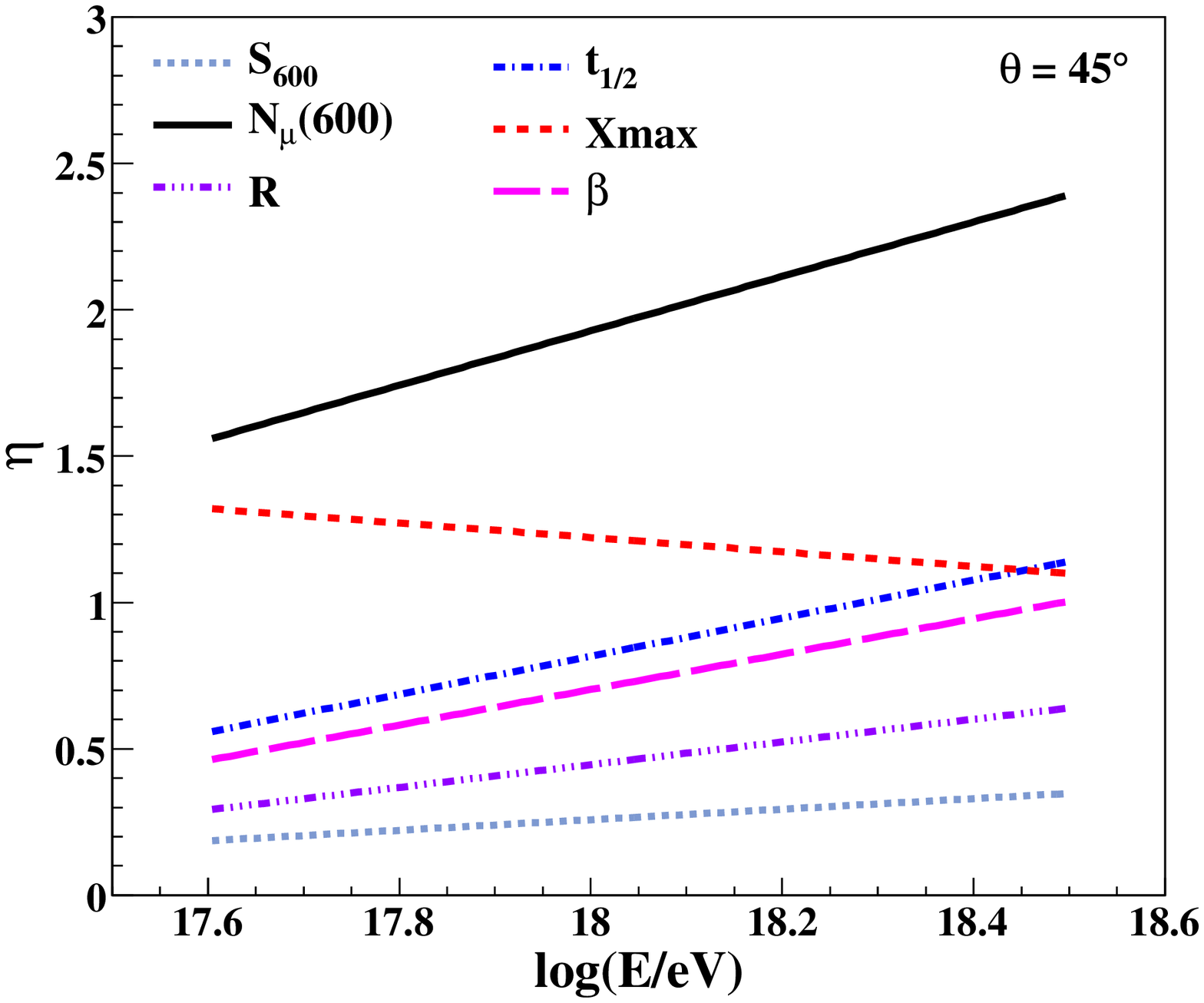}
\caption{Linear fits of $\eta$ as a function of $\log(E/\textrm{eV})$ for the
different parameters considered and for $\theta=30^\circ$ and $\theta=45^\circ$. \label{Etaplot}}
\end{center}
\end{figure}

Presumably, the interpolated signal of the water Cherenkov detectors 
at 600 m from the shower axis $S_{600}$ will be used to obtain 
the primary energy for the 750 m-array. Therefore, to study its dependence 
on the primary mass we included it in the set of parameters. Figure \ref{Etaplot} 
shows the linear fits of $\eta$ as a function of the logarithm of the energy 
for $\theta=30^\circ$ and $\theta=45^\circ$ and for the different parameters 
considered.

From figure \ref{Etaplot} we see that the parameter which better 
separates protons from iron nuclei appears to be $N_{\mu}^{Rec}(600)$, 
followed by $X_{max}$. We can also see that, except for $X_{max}$, 
$\eta$ increases with the primary energy. This happens because, as 
the energy increases, the number of triggered stations also increases, 
which reduces the reconstruction errors. The number of particles in the 
detectors also increases with the energy and therefore reduces 
fluctuations.

Although the reconstruction error of $X_{max}$ decreases with
primary energy, $\eta$ slowly decreases. This happens
because the mean values of $X_{max}$ corresponding to protons and
iron nuclei get closer as the energy increases for the hadronic model 
considered (however, this is not the case for the hadronic model EPOS, 
see Ref. \cite{Unger:07}).

\begin{figure}[!bt]
\begin{center}
\includegraphics[width=6.8cm]{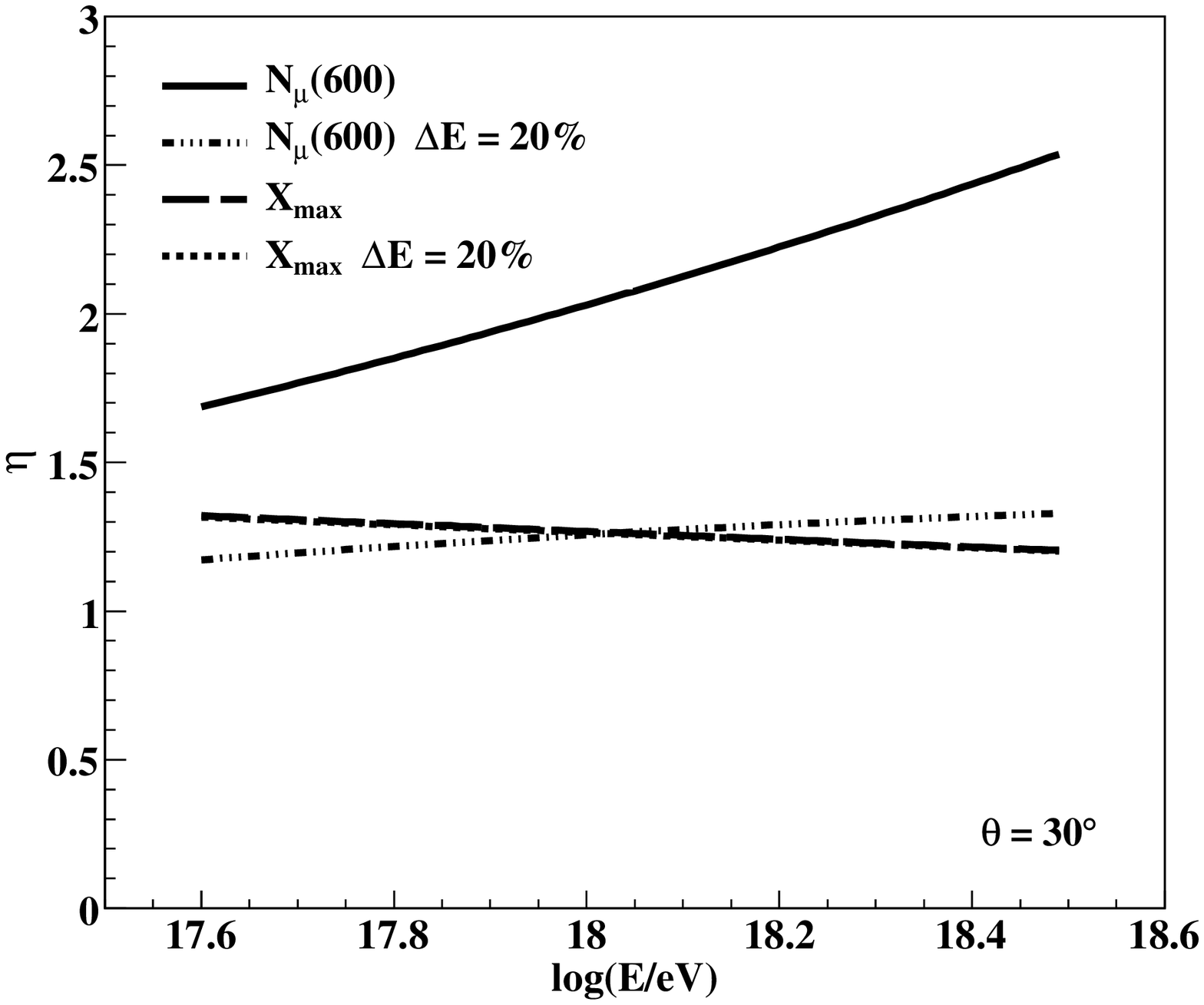}
\includegraphics[width=6.8cm]{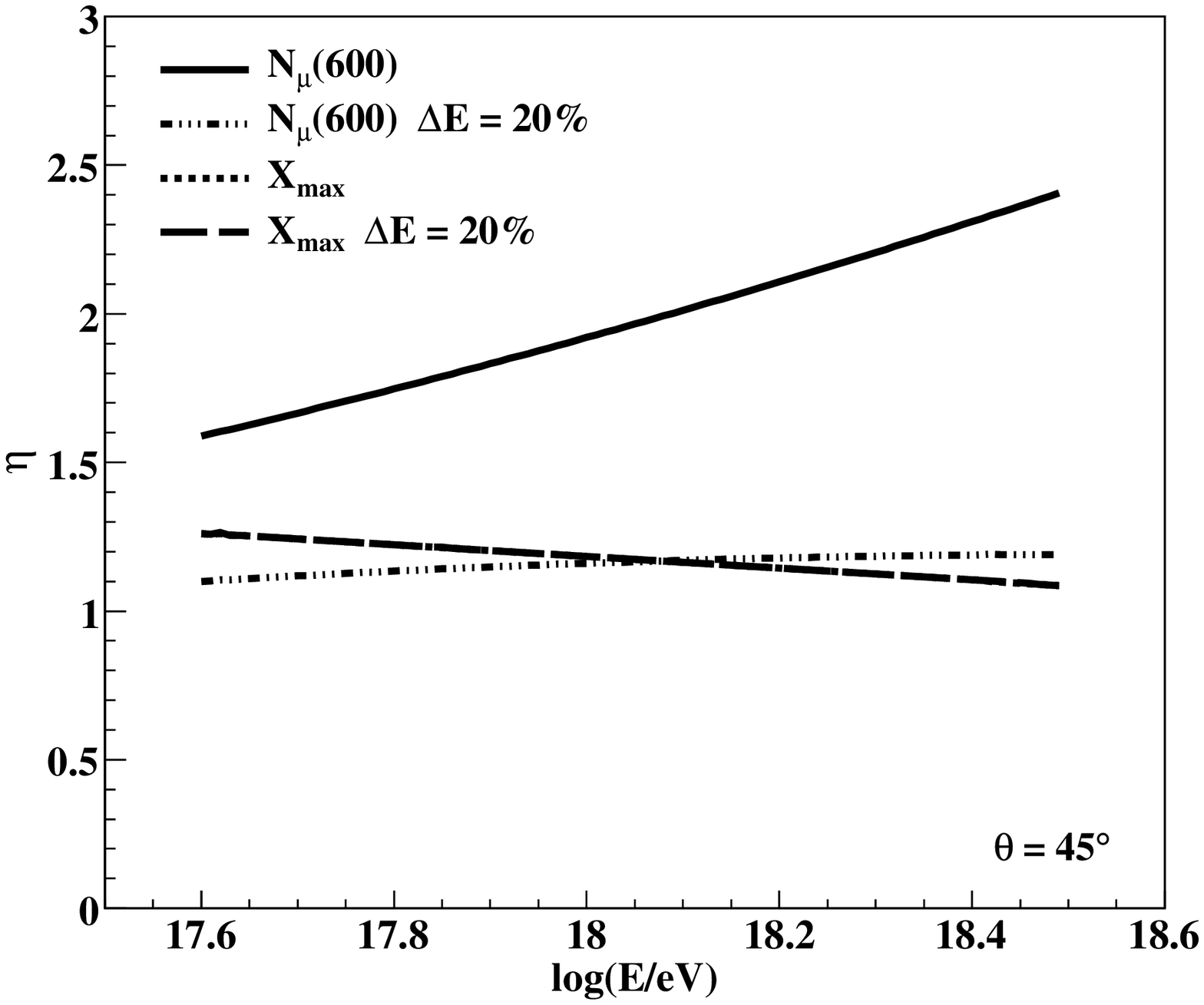}
\caption{$\eta$ as a function of $\log(E/\textrm{eV})$ for the parameters 
$X_{max}$ and $N_{\mu}^{Rec}(600)$ for $\theta=30^\circ$ and $\theta=45^\circ$,
assuming a power law spectrum of spectral index $\gamma=2.7$ and a gaussian 
uncertainty of $20\%$ of relative error and without including the effects of the
energy uncertainty (see appendix \ref{AppDE} for details). The differences between 
the curves corresponding to $X_{max}$ with and without including the energy uncertainty 
are so small that cannot be seen in both plots. \label{EtaplotDE}}
\end{center}
\end{figure}

If the energy uncertainty is negligible, $N_{\mu}^{Rec}(600)$ is the 
best parameter for mass discrimination analysis. However, as opposed 
to the other parameters which depend on the energy logarithmically, 
the number of muons is almost proportional to the primary energy. 
Therefore, the energy uncertainty will affect more the discrimination 
power of $N_{\mu}^{Rec}(600)$ than that of the other parameters. Figure
\ref{EtaplotDE} shows $\eta$ as a function of the energy assuming a 
$20\%$ energy uncertainty for the parameters $N_{\mu}^{Rec}(600)$
and $X_{max}$ (see appendix \ref{AppDE} for the details of the 
calculation). We see that the discrimination power of $N_{\mu}^{Rec}(600)$
decreases in such a way that it is of the order of the corresponding to
$X_{max}$. From figure \ref{EtaplotDE} we also see that the discrimination
power of $X_{max}$ remains approximately the same when the energy 
uncertainty is included, which is due to its logarithmic dependence
on the primary energy.

\section{Conclusions}

In this paper we studied the effect of muon counter
segmentation and we found an analytical expression to correct for
the undercounting due to muon pile-up. We presented a detailed
method to reconstruct the muon lateral distribution function for
the 750 m infill array. We also studied the potential of the 
$N_\mu^{Rec}(600)$ parameter as a discriminator of the identity of 
the primary cosmic ray particle.

We showed that for 192 segments and 90 incident muons in a 20 ns
time bin (maximum number of muons at 200 m from the shower axis
for $E = 10^{18}$ eV, which would correspond to saturation for 
the adopted configuration) the total error, Poissonian plus pile-up 
correction, does not exceed the Poissonian error by more than $14\%$. 
This correction is enough to attain, with surface information alone,
a composition discrimination power comparable, or even better, than 
that of $X_{max}$. Clearly, this result depends quantitatively on the 
assumed hadronic interaction model and, consequently, the safest 
conclusion is probably that a two-dimensional analysis, including both 
parameters, might well configure an optimal composition analysis method 
which will be discussed in an accompanying paper \cite{supa:07}.

The error in the determination of $N_\mu^{Rec}(600)$ parameter is 
$\lesssim 20\%$ at $10^{17.6}$ eV (energy threshold of the 750 m array) 
and decreases as the energy increases, being less than $12\%$ for 
$E \geq 10^{18}$ eV. So, at this energy or greater, this uncertainty is 
small compared with the expected primary energy uncertainty, $\sim 20\%$, 
which affects directly the determination of $N_\mu^{Rec}(600)$ since shower 
muon content is nearly proportional to the primary energy. Therefore, 
$30$m$^{2}$ individual muon detectors partitioned in 192 segments, can be 
considered as an optimal compromise between cost and scientific output.

We also studied the effect of the energy uncertainty on the discrimination 
power of $N_\mu^{Rec}(600)$. The inclusion of this effect somewhat degrades the 
discrimination power of $N_\mu^{Rec}(600)$, however, it generally 
remains as good a parameter as $X_{max}$.

\section{Acknowledgments}

The authors have greatly benefited from discussions with several colleagues 
from the Pierre Auger Collaboration, of which they are members. We also want to 
acknowledge Corbin Covault for carefully reading the manuscript and for his 
valuable comments. GMT acknowledges the support of DGAPA-UNAM through grant 
IN115707.

\appendix

\section{Effect of the Energy Uncertainty}
\label{AppDE}

We estimate the effect of the energy uncertainty on the discrimination power of
the parameters $N_\mu^{Rec}(600)$ and $X_{max}$, assuming a power law spectrum of spectral
index $\gamma = 2.7$, $J(E) \propto E^{-\gamma}$, and a gaussian uncertainty in the energy
determination of $\sigma = \varepsilon E$ with $\varepsilon$ the relative error. Let $q$ 
be a parameter for which we know its distribution function parametrized by the primary energy, 
$f(q;E)$. Therefore, the conditional probability of $q$ given the reconstructed energy $E_{rec}$ 
is,
\begin{equation}
P(q\ |E_{rec})= C(E_{rec})\ \int_{E_1}^{E_2} dE\ f(q;E)\ \exp\left( -\frac{(E_{rec}-E)^2}{2 E^2 \varepsilon^2} %
\right) E^{-\gamma-1},
\label{Pqe}
\end{equation}
where $E_1=10^{16}$ eV and $E_2=10^{20}$ eV are the lower and upper limits of the part of the spectrum 
considered, respectively, and,
\begin{equation}
C^{-1}(E_{rec}) = \int_{E_1}^{E_2} dE\ \exp\left( -\frac{(E_{rec}-E)^2}{2 E^2 \varepsilon^2} \right)%
E^{-\gamma-1}.
\label{C}
\end{equation}

From Eq. (\ref{Pqe}) we can calculate the mean value and the variance of the parameter $q$ as a 
function of the reconstructed energy,
\begin{eqnarray}
\label{MeanEu}
\langle q \rangle (E,\varepsilon) &=&  C(E_{rec})\ \int_{E_1}^{E_2} dE\ \langle q_0 \rangle (E)%
\ \exp\left( -\frac{(E_{rec}-E)^2}{2 E^2 \varepsilon^2} \right) E^{-\gamma-1}, \\  
Var[q](E,\varepsilon) &=& C(E_{rec})\ \int_{E_1}^{E_2} dE\ (Var[q_0](E)+\langle q_0 \rangle^2(E)) 
\times \nonumber \\
\label{VarEu}
&& \exp\left( -\frac{(E_{rec}-E)^2}{2 E^2 \varepsilon^2} \right) E^{-\gamma-1}
-\langle q \rangle^2 (E,\varepsilon),
\label{var}
\end{eqnarray}
where $\langle q_0 \rangle (E)$ and $Var[q_0](E)$ are the mean value  and the variance of parameter 
$q$ without taking into account the energy uncertainty, calculated using $f(q;E)$. Therefore, to calculate 
the parameter $\eta$ defined in Eq. (\ref{EtaDef}) including the energy uncertainty, we need the functions 
$\langle q_0 \rangle (E)$ and $Var[q_0](E)$ for each parameter considered. For that purpose, we fitted the 
mean value and the standard deviation of $N_\mu^{Rec}(600)$ with a function of the form $g(E)=a\ E^b$ and 
for $X_{max}$ with $h(E)=a+b \log(E)$. Using the fits obtained for $\theta=30^\circ$ and $\theta=45^\circ$ 
and Eqs. (\ref{MeanEu}), (\ref{VarEu}) and (\ref{EtaDef}) we obtain $\eta(E_{rec},\varepsilon)$ which can 
be seen in Figure \ref{EtaplotDE} for $\varepsilon=0.2$.

\end{document}